# Twin Universes and Bimetric Gravity

## From Sakharov's Cosmological Symmetry to Modern Speculations on Negative Mass

Jean-Pierre Luminet

*Laboratoire d'Astrophysique de Marseille (LAM), Centre National de la Recherche Scientifique, Aix-Marseille Université ;* jean-pierre.luminet@lam.fr

## Abstract

We review the conceptual and theoretical status of twin-universe models, from Sakharov's early ideas on baryogenesis, CPT symmetry, and reversed arrows of time to recent quantum-cosmological scenarios involving pair-created entangled universes. Particular attention is given to the distinction between Sakharov-type cosmological symmetry, modern bimetric theories of gravity, and speculative models involving negative mass sectors. We summarize the main consistency conditions faced by bimetric gravity, including the Boulware–Deser ghost, Higuchi-type bounds, covariant conservation laws, and no-go results for positive- and negative-mass sectors. We then examine the Janus cosmological model and related black-hole replacement proposals, emphasizing the difference between heuristic twin-universe narratives and fully specified relativistic field theories. We argue that while twin-universe ideas remain a fertile framework for questions about the arrow of time, matter–antimatter asymmetry, and the quantum origin of the universe, models combining bimetric gravity, negative mass, dark-sector phenomenology, and alternatives to Kerr black holes face a substantial burden of mathematical and observational proof.

## Introduction

A currently hotly debated issue in cosmology is the status of the standard model known as $\Lambda$CDM, which relies in particular (as its acronym indicates) on two components that ensure its consistency with all astronomical observations: non-baryonic cold dark matter (CDM) and dark energy ($\Lambda$) 1. Despite its effectiveness, however, it has been the subject of criticism, stemming in part from the fact that no non-baryonic dark matter particles have yet been detected experimentally, on the one hand, and that the existence of a repulsive—and thus antigravitational—form of dark energy relies on certain extrapolations from quantum field theory, on the other. This is why numerous attempts have been made to develop alternative theories to the Standard Model that would eliminate the need for dark matter and/or dark energy. These attempts are based on various extensions—or even radical modifications—of Einstein's theory of general relativity, upon which the current Standard Model is built 2. One such attempt, long studied by many researchers, has ultimately failed: MOND (Modified Newtonian Dynamics) theories 3.

As a result, other, more profound approaches have recently begun to emerge, rooted in part in a fascinating concept proposed as early as in the 1960's by the Russian physicist Andrei Sakharov: twin universes 4.

## The Ambiguous Power of Twin Universes

Few cosmological ideas are as suggestive as that of a twin universe. After Sakharov it has reappeared many times in modern cosmology, being invoked to address the arrow of time, the matter–antimatter asymmetry, the initial quantum state of the universe, dark matter, dark energy, and even the nature of black holes. Yet the same expression—"twin universe"—often denotes very different concepts: a CPT-conjugate cosmological branch, a mirror sector of matter, a pair-created quantum universe, an entangled partner universe, or a second metric sheet interacting with our own.

This semantic richness is also a danger. A twin universe in the sense of Sakharov is not automatically a bimetric universe. A CPT-conjugate branch is not automatically a world of negative mass. A mirror sector is not automatically a substitute for dark matter or dark energy. A second metric is not automatically a coherent theory of gravity. Much of the confusion surrounding modern twin-universe cosmologies comes from sliding too easily between these meanings.

The aim of this paper is not to dismiss the twin-universe idea. On the contrary, its history shows that it has generated serious physics. Sakharov used it in connection with the matter–antimatter asymmetry and the cosmological arrow of time. Recent quantum-cosmological work has used pair-created universes and entanglement to formulate possible observational signatures in the cosmic microwave background. Bimetric gravity, meanwhile, has become a technically sophisticated field after the development of ghost-free massive gravity and the Hassan–Rosen theory 5. Our paper also examines attempts to introduce negative-mass sectors, including the no-go theorem of Hohmann and Wohlfarth 6 and Hossenfelder's response 7.

The difficulty begins when these distinct traditions are merged too quickly. This is the case with the Janus cosmological model 8, 9 and related "plugstar" proposals 10. A model that invokes Sakharov, two metrics, negative masses, cosmic acceleration, large voids, antimatter, and alternatives to black holes may be imaginative; but imagination is not yet a theory. A theory must specify its action, its degrees of freedom, its conservation laws, its weak-field limit, its perturbative stability, and its observational predictions. This distinction between image and theory will guide our discussion.

The latter part of this article will focus on the shortcomings of the Janus model, without, however, going into too much technical detail. The general conclusion is that Sakharov's twin-universe intuition remains fertile, and bimetric gravity remains a legitimate but highly constrained field, whereas Janus-like models combining twin universes, negative mass, dark-sector phenomenology, and black-hole replacements face a heavy burden of mathematical and observational proof.

## Sakharov: Baryogenesis, CPT, and the Arrow of Time

The modern history begins with Andrei Sakharov's 1967 work on the origin of the matter–antimatter asymmetry 4. Sakharov formulated the now famous conditions for baryogenesis: violation of baryon number, violation of C and CP symmetries, and departure from thermal equilibrium. These conditions addressed a simple but profound question: if the early universe was governed by laws that treat matter and antimatter almost symmetrically, why does the observable universe contain overwhelmingly more matter than antimatter?

Sakharov's later cosmological reflections extended this concern with symmetry. He considered models in which the universe may possess branches with opposite arrows of time and discussed the hypothesis of cosmological CPT symmetry 11. His work explicitly mentions cosmological models with a reversal of the time arrow, a cosmological CPT-symmetry hypothesis, and many-sheeted open models 12.

The key point is often missed. Sakharov's twin-universe idea was not a theory of negative gravitational mass. Nor was it a bimetric theory in the modern sense. It was first a proposal about global symmetry: perhaps the universe as a whole restores a symmetry that appears broken in our local branch. The arrow of time may point in opposite directions in two cosmological domains, but observers within each domain would still experience entropy increasing toward their own future.

This is a subtle idea. It does not mean that inhabitants of the twin branch would see broken cups reassemble or stars unburn. It means that the global structure relating the two branches may be time-reversed, while each branch has its own thermodynamic orientation.

Sakharov's legacy is therefore real but specific. He opened the possibility of a cosmological doubling associated with CPT, baryogenesis, and the arrow of time. However Sakharov did not provide a license to identify the second branch with a negative-mass universe, nor to infer that black holes must be replaced by transitions into another sheet.

In 2018, Boyle, Finn and Turok 13 published a cosmological model based on the existence of a mirror universe to our own, populated by antimatter and "running backward in time," similar to Sakharov's model in the context of CPT symmetry. Their explanation has been highlighted as "obvious" for the presence of matter and the absence of antimatter in our known universe. The authors introduced a new hypothetical species of very massive neutrino (more than 500 million times heavier than the proton, or $4.8\times10^{8}$ GeV/$c^{2}$) as a candidate explanation for the nature of dark matter.

## Pair-Created and Entangled Universes: The Quantum-Cosmological Line

The most natural continuation of Sakharov's intuition is not necessarily bimetric gravity. It is quantum cosmology. In that setting, the universe is treated not merely as a classical space-time solving Einstein's equations, but as an object described by a quantum state, an instanton, or a gravitational path integral.

This is where recent work by Chen, Lin, Lin and Yeom is especially relevant 14. Their paper asks whether the origin of the universe and the arrow of time might be related. The authors consider the possibility that our universe was pair-created with a twin whose arrow of time is opposite to ours. They argue that such twin universes should naturally be entangled, and that the entanglement may leave signatures in the cosmic microwave background.

Their framework belongs to Euclidean quantum gravity. In their model, the two universes are bridged by a Euclidean wormhole, specifically a Klebanov–Susskind–Banks-Fishler wormhole used as a tractable toy model 15. Each universe is then described as a mixed state when considered alone, because the full state belongs to the pair. The entanglement selects a global vacuum for the inflaton perturbations, and the authors find that this can enhance the CMB power spectrum for long-wavelength modes. They describe this enhancement as a possible "birthmark" of the pair-creation process.

The importance of this work is not that it establishes the existence of a twin universe. Its importance is methodological. It formulates a specific mechanism, derives a specific class of effects, and points to a possible observational signature. The claim is not that one can explain everything—dark matter, dark energy, antimatter, voids, black holes—by invoking a twin universe. The claim is narrower and therefore more scientifically valuable: if a pair-created entangled twin existed, it might leave an imprint at large angular scales in the CMB spectrum.

This distinguishes the Chen–Lin–Lin–Yeom approach from more expansive cosmological narratives. It remains speculative, but the speculation is disciplined. It is connected to a mechanism, a calculation, and a possible test.

## From Twin Universes to Bimetric Gravity: A Change of Register

To move from Sakharov or Chen et al. to bimetric gravity is to change the nature of the problem. A twin universe may be a CPT-conjugate branch, a quantum partner, or a mirror sector. Bimetric gravity is something more specific: it introduces two metric tensors, often denoted $g_{\mu\nu}$ and $f_{\mu\nu}$, each capable of defining a geometry.

In general relativity, the metric is not merely a measuring device. It defines distances and times, but also the causal structure, the Levi-Civita connection, geodesics, curvature, and the covariant conservation law for the stress-energy tensor. Introducing a second metric therefore introduces a second geometry, a second connection, and potentially a second set of geodesics.

This is why bimetric gravity is not obtained by simply writing down two Einstein equations and coupling them by intuition. The Bianchi identities impose severe constraints. For a metric $g_{\mu\nu}$, the Einstein tensor $G^{\mu\nu}$ satisfies

$$\nabla^{(g)}_{\mu} G^{\mu\nu}(g) = 0.$$

If matter is coupled to $g_{\mu\nu}$, consistency normally requires

$$\nabla^{(g)}_{\mu} T^{\mu\nu} = 0,$$

where $T^{\mu\nu}$ is the stress-energy tensor. With two metrics, one must specify which covariant derivative acts on which stress tensor, and how interaction terms between the two sectors satisfy the corresponding constraints. A tensor may be covariantly conserved with respect to one connection and not with respect to the other. This technical point is not a detail; it is the backbone of the theory.

A second metric also changes the spectrum of gravitational degrees of freedom. In a healthy bimetric theory one typically expects a massless spin-2 field and a massive spin-2 field. A massless spin-2 field has two physical polarizations. A massive spin-2 field has five. The danger is that a sixth, pathological degree of freedom may appear. This is the Boulware–Deser ghost 16.

## The Boulware–Deser Ghost and the Modern Revival of Bimetric Gravity

The Boulware–Deser ghost was one of the main reasons nonlinear massive gravity and bimetric gravity were long regarded with suspicion. At the linear level, the Fierz–Pauli theory 17 can describe a massive spin-2 field without immediately generating the unwanted sixth mode. But generic nonlinear extensions reintroduce it. The problem is not merely aesthetic. A ghost corresponds, roughly speaking, to a degree of freedom with the wrong sign of kinetic energy, threatening catastrophic instability of the vacuum.

Modern massive gravity was revived by the de Rham–Gabadadze–Tolley construction 18, which identified a special class of interaction terms capable of avoiding the ghost. Hassan and Rosen 5 then extended this achievement to bimetric gravity by making the reference metric dynamical. Their paper states the point clearly: generic nonlinear bimetric theories suffer from the same Boulware–Deser ghost instability as nonlinear massive gravity, but a special construction yields a nonlinear bimetric theory describing a massless spin-2 field interacting with a massive spin-2 field and free of that ghost.

The structure of the Hassan–Rosen theory is highly constrained. Each metric has its own Einstein–Hilbert term, and the interaction potential is built from elementary symmetric polynomials of the square-root matrix

$$\sqrt{g^{-1}f}$$

This square root is not a decorative mathematical object. It is part of the precise algebraic structure required to maintain the constraint that removes the ghost.

The lesson is important for any twin-universe or two-sheet model that claims to be bimetric. A bimetric theory is not rendered viable by the mere presence of two metrics. It must show that its degrees of freedom are controlled, that its interaction terms are consistent, and that it does not propagate an unstable ghost.

## The Higuchi Bound and Cosmological Stability

Even when the Boulware–Deser ghost is avoided, a bimetric or massive-gravity model can fail in cosmology. A theory may have acceptable background equations and still become unstable at the level of perturbations.

One classical warning is the Higuchi bound 19. In de Sitter space, a massive spin-2 field cannot have an arbitrary mass. Higuchi showed that, for spin 2 in the expanding part of de Sitter space, there is a forbidden mass range associated with negative-norm states. In the usual notation, the bound is often written as

$$m^2 \geq 2H^2,$$

where $m$ is the spin-2 mass and $H$ is the de Sitter expansion rate. The exact form of the constraint may vary in generalized bimetric settings, but the physical message remains: the helicity-zero mode of a massive graviton can become pathological if the mass scale is too small compared with the curvature scale.

This matters because many bimetric cosmologies are attractive precisely at large scales. They may mimic dark energy or modify the expansion history. But if the perturbations are unstable, the background success is illusory.

The same warning applies to couplings with matter. In many ghost-free constructions, ordinary matter is coupled to only one metric. Coupling matter simultaneously to both metrics, or introducing two matter sectors interacting through both metrics, can reintroduce instabilities. Thus the apparently natural idea that each universe sheet has its own matter and its own metric is not automatically safe.

The present status of bimetric gravity is therefore balanced. It is a legitimate theoretical field, not a fringe speculation. But it is not an established replacement for general relativity or $\Lambda$CDM. Its viability depends on a restricted parameter space, stability conditions, matter couplings, and confrontation with cosmological and gravitational-wave data.

## Negative Mass and the No-Go Problem

The idea of negative mass had been introduced in 1957 by Hermann Bondi in the very specific context of testing the conceptual limits of general relativity and the equivalence principle 20. Bondi was not proposing negative mass as an observed component of the Universe, nor primarily as a cosmological model. His question was more foundational: does general relativity itself logically forbid matter whose mass has the opposite sign? His conclusion was essentially that negative mass is not excluded by a simple local contradiction in the equations, although it leads to deeply pathological and physically implausible behaviour. However the concept of negative mass has been introduced in twin-universe models because it seems to offer a simple explanation of repulsion 21. Positive masses attract; perhaps negative masses repel; perhaps a negative-mass sector could carve out cosmic voids or mimic dark energy. The intuition is tempting but treacherous.

One must distinguish inertial mass, passive gravitational mass, and active gravitational mass. In Newtonian language, these distinctions can be manipulated by signs. In relativistic language, however, the source of gravity is the stress-energy tensor, and its conservation is tied to the geometry. “Negative mass” must therefore mean something precise: negative energy density? negative active gravitational source? negative passive response? a sector coupled to a different metric? an effective sign reversal produced by a nonstandard interaction?

Without these distinctions, the concept becomes ambiguous.

The classical problem is runaway motion. In a naive model with positive and negative masses, a pair can self-accelerate in a way that appears to violate ordinary energy conditions. Bimetric models have sometimes been proposed to avoid this by assigning each sector its own metric and geodesics. But this does not solve the problem by itself. The weak-field limit must be derived from covariant equations.

This is the significance of the Hohmann–Wohlfarth no-go theorem 6. They considered what they described as the most conservative geometric extension of Einstein gravity containing positive and negative mass sources and observers: a bimetric theory with two copies of standard-model matter interacting only gravitationally. They then studied the most general linearized field equations satisfying physically and mathematically motivated assumptions and proved a no-go theorem: in such a framework, one cannot obtain precisely opposite Newtonian forces on positive and negative test masses.

A no-go theorem does not prove that all negative-mass theories are impossible. It proves that a desired result cannot be obtained under specified assumptions. A model may escape the theorem by abandoning one of those assumptions. But this is not a free escape.

If a model abandons a standard weak-field limit, a conservation law, a coupling principle, or a symmetry between sectors, it must show what replaces it and why the replacement is physically acceptable.

Hossenfelder's work is relevant here. She proposed a bimetric theory with exchange symmetry 22 and later argued that the Hohmann–Wohlfarth no-go theorem did not apply to her model 7. Her response does not establish a complete viable cosmology of negative masses, but it does show the proper level of discussion: one must formulate the theory covariantly, specify the metrics and connections, and identify which assumptions of a no-go theorem are not satisfied.

This is an important distinction. Negative mass is not forbidden by a slogan. But it is not rescued by a slogan either.

## The Janus Model: Ambition and Burden of Proof

The Janus cosmological model 9, 23 occupies a distinctive place in this landscape. It explicitly claims descent from Sakharov's twin-universe idea and proposes a bimetric cosmology with two interacting universe sheets. The authors describe the standard $\Lambda$CDM model as facing challenges involving vast cosmic voids, early formation of stars and galaxies, and matter–antimatter asymmetry; they then propose interactions between two universe sheets through a bimetric model as an alternative interpretation of large-scale structure, voids, and cosmic acceleration. The publication of such a model in peer-reviewed journals makes it worth examining but it does not, by itself, validate all of the broader claims associated with the Janus program.

The first issue is the appeal to Sakharov. The historical connection is understandable: Sakharov did consider twin universes, reversed arrows of time, and large-scale symmetry. But Janus adds several strong hypotheses not contained in Sakharov's original idea. In particular, it introduces negative-mass sectors and bimetric interactions as explanatory mechanisms for cosmic structure and acceleration. These may be proposed as extensions; they should not be presented as direct consequences.

The second issue is the status of the bimetric structure. If Janus is a bimetric theory, then it must be judged by the standards appropriate to bimetric theories. Does it possess an action principle? Are the two metrics dynamical? What are the interaction terms? How many degrees of freedom propagate? Is the Boulware–Deser ghost absent? Are the perturbations stable? Are the Bianchi identities satisfied with the appropriate covariant derivatives?

Unfortunately these questions are not addressed in a fully satisfactory way. Their model's equations are

$$G_{\mu\nu}(g) = \chi\left(T_{\mu\nu} + K_{\mu\nu}\right) \text{ and } \bar{G}_{\mu\nu}(\bar{g}) = \kappa\bar{\chi}\left(\bar{T}_{\mu\nu} + \bar{K}_{\mu\nu}\right)$$

where $K_{\mu\nu}$ is the interaction tensor representing the contribution to the field acting on positive masses, arising from the presence of negative masses, and $\overline{K}_{\mu\nu}$ is the interaction tensor representing the contribution to the field acting on negative masses, arising from the presence of positive masses.

Such a system must satisfy Bianchi's conditions. The authors distinguish between two cases: a non-stationary homogeneous-isotropic case and a stationary case with *O(3)* symmetry, starting with Schwarzschild. They then invoke, for the cosmological case, specific forms $K_{\mu\nu} = \varphi(t)\overline{T}_{\mu\nu}$ and $\overline{K}_{\mu\nu} = \phi(t)T_{\mu\nu}$, leading to generalised energy conservation. For the steady-state case, their own formulation refers to the compatibility of the Bianchi conditions in the Newtonian approximation. Thus the published Janus equations are arranged so that Bianchi-type consistency conditions are only verified in specific sectors, through *ad hoc* or specialised choices of interaction tensors, without a general proof of the covariant consistency of the full bimetric system.

The third issue is the weak-field limit. If the model relies on attraction within one sector and repulsion between sectors, those effects must be derived from the covariant equations, not imposed through Newtonian intuition. In light of the Hohmann–Wohlfarth theorem, the model must also specify which assumptions it abandons and how it avoids the corresponding difficulties. This is not rigorously the case with Janus. In order to ensure that the Newtonian approximation in the weak-field limit, the authors state that in stationary conditions, the square root of the ratio of the determinants are nearly constant: $\sqrt{|\bar{g}|/|g|} \approx 1$. This is an unchecked assumption: even in the stationary case, the metric determinant can depend on the radial coordinate.

The fourth issue is observational. A model may offer qualitative interpretations of cosmic voids, early structures, or acceleration. But modern cosmology is not constrained by one observation. It is constrained by an interlocking network: CMB anisotropies, baryon acoustic oscillations, supernovae, primordial nucleosynthesis, weak and strong lensing, cluster abundances, galaxy clustering, gravitational waves, and local tests of gravity. A serious alternative must confront this network quantitatively.

This is the standard required of any model that aims to replace or substantially modify $\Lambda$CDM plus general relativity.

## Plugstars, Schwarzschild, and the Missing Kerr Problem

The most delicate claims associated with the Janus program concern black holes. In some presentations 10, 24, black holes are said to be mathematically inconsistent or physically replaced by objects sometimes called "plugstars." Matter falling into such an object is described as undergoing a transition into another sector or sheet, sometimes with a change of mass sign.

Here one must separate three issues.

First, the authors repeat an error originally made in 1999 by the first English-language translators of Schwarzschild's original article (in German) 25. According to them, the metric initially derived by Schwarzschild, written in a particular coordinate system, is not at all equivalent to the "standard" metric established later by David Hilbert, as used in the classic monographs. In the former, the formation of an event horizon—and thus a black hole—would be ruled out, but this result would have been ignored by the community of theorists, who would have simply repeated Hilbert's "error" for a century without verifying their German-language source!

This claim was quickly refuted by several authors 26, 27, who demonstrated that the original Schwarzschild solution is entirely equivalent to the solution established by Hilbert.

Second, the Kruskal's maximally extended Schwarzschild solution 28 is an idealized eternal vacuum solution. It is mathematically rich, but it is not the space-time of a realistic gravitational collapse. The Kruskal extension contains regions that do not arise in the same way in a collapse geometry. One cannot infer, merely from the maximal Schwarzschild extension, that matter in a real astrophysical collapse passes into another universe or changes its mass sign.

Third, even if one focused only on Schwarzschild, criticism of an idealized nonrotating solution would not suffice to invalidate astrophysical black holes. Real compact objects form through collapse, accretion, and mergers. Angular momentum is generically present.

The relevant stationary solution for an astrophysical rotating black hole is Kerr, not Schwarzschild. Kerr space-time 29 is characterized by mass ($M$) and angular momentum ($J$). Its physics includes an ergosphere, frame dragging, photon rings, innermost stable circular orbits, relativistic accretion disks, jets, and quasi-normal modes. These are not optional details; they dominate the phenomenology of real black-hole candidates.

The Event Horizon Telescope observations of M87* and Sgr A* are interpreted within precisely this framework. The EHT analysis of M87* 30 found the observed asymmetric ring consistent with strong gravitational lensing around a spinning Kerr black hole, and it emphasized the role of black-hole spin in models producing powerful jets. The EHT results for Sgr A* 31 likewise report consistency with the expected appearance of a Kerr black hole.

This makes the absence, or near absence, of a serious Kerr analysis in black-hole replacement proposals a major weakness. A model that claims to replace black holes cannot restrict itself to Schwarzschild. It must reproduce or replace the phenomenology of Kerr: ring diameter and asymmetry, lensing structure, frame dragging, accretion signatures, jet formation, polarization, gravitational-wave ringdown, and the dynamics of compact-object mergers.

A critique of Schwarzschild alone misses the main target. Astrophysical black holes are not static pedagogical spheres; they are rotating relativistic engines. Any proposed alternative must meet them where the data are.

## What Remains Valuable

The twin-universe idea remains valuable for several reasons.

It preserves a deep question about symmetry. The universe we observe is not manifestly symmetric between matter and antimatter, nor between the two directions of time. Sakharov's insight was to ask whether a larger cosmological structure might restore a symmetry hidden from our branch.

Next, it gives quantum cosmology a natural arena. Pair-created universes, entangled branches, and mixed reduced states are speculative ideas, but they are not merely literary images. They belong to attempts to describe the origin of the universe using quantum principles. The Chen et al. 14 proposal is particularly interesting because it connects the idea of an entangled twin to a possible CMB signature.

Eventually, it reminds us that the standard model of cosmology is incomplete. The nature of dark matter and dark energy remains unknown; the initial condition of the universe is mysterious; the arrow of time is not fully understood; and tensions in cosmological parameters motivate scrutiny. These gaps justify theoretical boldness.

But boldness must not be confused with consolidation. A model that explains many things at once is not automatically stronger than one that explains few. It is often more vulnerable, because each additional explanatory claim adds constraints.

A theory combining twin universes, two metrics, negative mass, cosmic acceleration, voids, antimatter, and black-hole alternatives must satisfy all the constraints relevant to each ingredient. It must be coherent as a relativistic field theory, stable as a perturbative theory, accurate as a cosmological model, and successful as an astrophysical model of compact objects.

## Conclusion

The history of twin universes is a history of fertile ambiguity. Sakharov's version belongs to the physics of symmetry, baryogenesis, and the cosmological arrow of time. Modern quantum-cosmological versions, such as the work of Chen and collaborators, explore whether pair-created entangled universes could leave observable imprints in the CMB. Bimetric gravity, in the Hassan–Rosen tradition, is a technically serious extension of general relativity, but one whose consistency depends on highly constrained interaction terms and stability conditions.

These three lines of thought should not be conflated. Sakharov does not automatically imply bimetric gravity. Bimetric gravity does not automatically imply negative mass. Negative mass does not automatically explain dark energy or cosmic voids. And none of these ideas automatically abolishes black holes.

The Janus model is best understood as an ambitious attempt to combine several of these strands. Its ambition makes it interesting, but also exposes it to a high burden of proof. It must show not only that its imagery is compelling, but that its equations are covariant, its conservation laws consistent, its degrees of freedom healthy, its weak-field limit correct, its perturbations stable, and its observational predictions competitive with existing models. The essential distinction is between a narrative and a theory. A narrative connects ideas. A theory connects equations. A narrative can suggest that a second universe restores cosmic symmetry. A theory must specify what fields exist, how they interact, what observers measure, and which data could falsify it.

Twin universes remain a powerful cosmological idea. They may yet illuminate the arrow of time, the initial state of the universe, or the origin of the matter–antimatter asymmetry. But their scientific future depends on discipline as much as imagination. The challenge is not to multiply cosmic mirrors, but to make one of them reflect testable physics.

Acknowledgement : The author thanks Jean-Pierre Petit for criticism and discussion.